\documentstyle[aaspp4,11pt,psfig]{article}
\newcommand{\etal}{{\it et al.\ }}

\begin{document}

\title{The millimeter VLBI Properties of EGRET Blazars}
\author{Geoffrey C. Bower \\
NRAO-Socorro and MPIfR \\
}

\begin{abstract}
We give a progress report on a snapshot 86 GHz-VLBI survey of the EGRET blazars
with the observatories of the CMVA.  A high fraction  (17/18) of the 
EGRET blazars were detected on the Pico~Veleta-Onsala baseline with a 
baseline length on the order of 500 $M\lambda$.
The detection threshold on the Pico~Veleta-Onsala baseline was $\sim 0.2$~Jy.
Six of these sources were not previously detected with 
3-millimeter VLBI.  We also present the detection of three new
non-EGRET sources.  The high detection rate for EGRET sources indicates
that gamma-ray flux is a robust predictor of millimeter wavelength
intensity.  Future more sensitive high-energy gamma-ray experiments
should find a larger class of objects detectable with millimeter
wavelength VLBI.
\end{abstract}

\section{Introduction}

A high fraction of high energy gamma-ray sources detected with the
EGRET telescope are known to be associated with blazars (Mattox \etal
1997).  Many of these blazars are among the brightest and most
variable sources at millimeter wavelengths.  Surveys at 
centimeter wavelengths have shown that a high fraction of the
EGRET blazars are detected on long baselines with VLBI (Moellenbrock
\etal 1996, Kellerman \etal 1998).  The conditions necessary to
produce high energy gamma-rays are similar to those for compact
radio sources.

For these reasons, it is likely that many of the EGRET blazars are
also millimeter wavelength VLBI sources.  Previous source compilations
and surveys indicate that the EGRET blazars comprise between
1/3 and 2/3 of all 86 GHz-VLBI detections (Rogers 1994, Lonsdale,
Doeleman \etal 1998).  However, a survey of these objects has 
not been performed with high sensitivity and under good weather
conditions.  

We report here on the first epoch of our survey of the EGRET
blazars.  The full survey will observe all EGRET blazars in
the Mattox \etal (1997) sample with $\delta > -20^\circ$.
All of these sources have $S_5 \ga 1 {\rm\ Jy}$ and flat
spectral indices ($\alpha \ga -0.5$).  Additional sources were
observed in the program.  These were selected from the Moellenbrock \etal
(1996) and Kellerman \etal (1998) surveys as sources with long
baseline fluxes greater than $\sim 1$~Jy.

\section{Observations and Results}

Observations were made on 2 April 1998 between 0 and 8 UT with
a global array of millimeter telescopes.  Three sources were observed
per hour, each for 6.5 minutes.  Additional time was given for
pointing and flux density measurements.
Due to poor weather and to equipment failures, reliable data was
obtained for only 3 stations:  Pico~Veleta, Bonn and Onsala.
Coverage in the visibility plane was essentially linear.
Data were correlated at Haystack and analyzed with the 
Haystack Observatory Post-processing Software (HOPS).

Standard polynomial gain curves were applied for Onsala and Bonn.
Antenna temperature measurements on source were used to calibrate Pico
Veleta.  These antenna temperature measurements were also used
to determine the zero-baseline flux of the sources.

We summarize in Table~1 the results of detection and model-fitting.
Signal-to-noise ratios for the incoherent average
are reported.  A snr greater than 3.5 is considered a detection.  
Inspection of the fringe rate and delay solutions for 1739+522 indicates
that the detection at $s_{SX}=3.8$ is firm.  Similarly, MK 421 ($s_{SX}=2.8$)
was clearly not detected.  The sources 0954+658 and 1219+285 were
not observed by Pico~Veleta.

Model-fitting was performed in two ways.  For sources detected on all
three baselines we fit a single Gaussian component to the three visibilities.
For sources not detected on the Bonn-Onsala baseline, we fit a single
component Gaussian to the zero baseline flux and the Pico~Veleta-Onsala
visibility.  We show in Figure~1 the results for the source 1606+106.

\begin{figure}[tb]
\begin{center}
\caption{Amplitude as a function of baseline length for the EGRET
blazar 1606+106.  The solid line shows the model Gaussian fit
to the three visibility points.  The zero-baseline flux is also
shown.  This source was detected with 3 mm VLBI for the first time
in this survey.}
\mbox{\psfig{figure=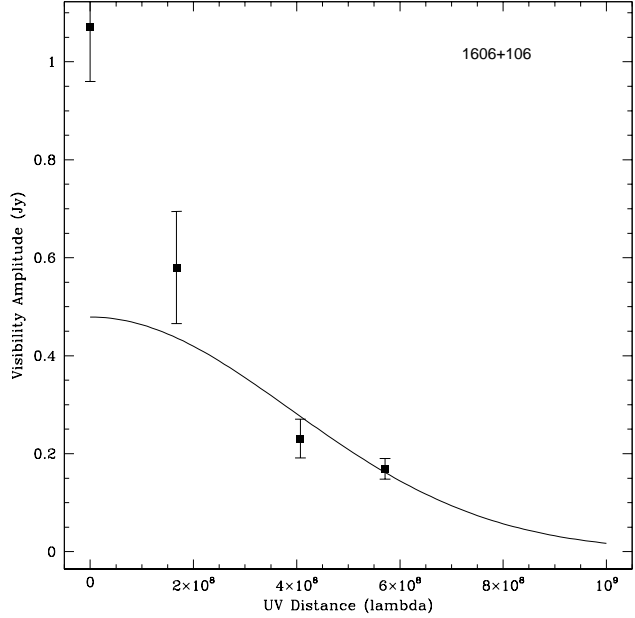,width=0.5\textwidth}}
\end{center}
\end{figure}

The errors given are the formal error assuming thermal
noise and a 10\% amplitude calibration error.  This significantly
underestimates the error for two reasons.  First, the source
structure may be more complex than a single component.  This leads
us in most cases to overestimate the size and underestimate the
brightness temperature.  However, in the event of beating between
two components, we may underestimate the size significantly.  Second,
calibration errors may in fact be much larger than 10\%.  The 
low correlated fluxes on the baselines to Bonn for 3C 454.3, BL Lac
and CTA 102 are suggestive of a pointing error.  These three
sources were observed in succession and may have been affected by
poor weather.

We do note that for many sources the Gaussian model indicates that
most of the zero-baseline flux is recovered on the short baselines.  
However, in none of the sources is the zero-baseline flux fully
recovered on the long baselines.  That is, all sources are resolved
to some extent on baseline lengths of 500 $M\lambda$.

Closure phases are also included in Table~1.  The sources 3C 279
and 3C 454.3 show significantly non-zero closure phases.

\section{Discussion}

A high fraction (17/18) of the EGRET blazars observed on the 
Pico~Veleta-Onsala baseline were detected.  Six of the
sources were detected for the first time.  The detection threshold on
this baselines was $\sim 0.2$~Jy, making this the most
sensitive 3-mm VLBI survey to be performed.  Many sources with total
fluxes below 1 Jy were detected.  We also detected four non-EGRET
blazars, three of them for the first time.

The high fraction of EGRET blazars detected supports the
conclusion drawn at lower frequencies that peak gamma-ray intensity
is a strong predictor of millimeter wavelength intensity.
This has several implications.

One, currently unidentified EGRET sources may be associated with
specific objects through high frequency VLBI surveys of sources
within the error box.  The Third EGRET catalog contains 170
unknown sources (Hartman \etal 1999).  The principal difficulty
here will be discriminating the compact non-EGRET blazar sources
from the compact EGRET sources.

Two, improvements in gamma-ray telescope sensitivity will produce
a much larger class of source available for study.  The GLAST
mission will be 30 times more sensitive than EGRET, implying a possible
increase in the source counts by a factor of $\sim 150$.
These sources will be of intrinsic interest, of use as a
phase and flux calibrators for other areas of research and of use as
probes of the intervening molecular gas.

Three, in order for these sources to be accessible to millimeter
VLBI, array sensitivity must be improved.  The results presented
here on the Haystack water vapor radiometer are very encouraging
in this regard (Tahmoush \& Rogers 1999, these proceedings).

\begin{deluxetable}{lrrrrrrrc}
\tablecaption{Blazar Detections at 86 GHz}
\tablehead{
\colhead{Source} & \colhead{I\tablenotemark{a}} & \colhead{$s_{BS}$\tablenotemark{b}} & \colhead{$s_{BX}$\tablenotemark{c}} &
\colhead{$s_{SX}$\tablenotemark{d}} & \colhead{$S_0$\tablenotemark{e}} & \colhead{$\theta$\tablenotemark{f}} & \colhead{$\phi$\tablenotemark{g}}  &\colhead{New?} \\
                 & \colhead{(Jy)} &                           &                     &
                   & \colhead{(Jy)} &  \colhead{(mas)}             &   \colhead{(deg)} 
}
\tablenotetext{a}{Zero-baseline flux.}
\tablenotetext{b}{Incoherently-averaged SNR for Bonn-Onsala baseline.}
\tablenotetext{c}{Incoherently-averaged SNR for Bonn-Pico Veleta baseline.}
\tablenotetext{d}{Incoherently-averaged SNR for Pico Veleta-Onsala baseline.}
\tablenotetext{e}{Total flux density of fitted Gaussian component.}
\tablenotetext{f}{Circular FWHM of fitted Gaussian component.}
\tablenotetext{g}{Closure phase.}
\startdata
\multicolumn{9}{c}{EGRET sources} \\
\hline
0716+714 &  1.74 &   6.8 &  24.4 &  26.5 & $  1.27 \pm  0.13 $ & $  0.011 \pm  0.075 $ & $-7.8 \pm 2.9$ \\
0827+243 &  2.19 &   3.4 &  13.6 &  19.1 & $  1.10 \pm  0.12 $ & $  0.008 \pm  0.062 $ & $11.4 \pm 6.5$ \\
0917+449 &  0.80 &   1.9 &   2.8 &   9.3 & $  0.80 \pm  0.20 $ & $  0.135 \pm  0.010 $ & \dots & $\surd$ \\
0954+658 & 0.30  &   3.2 &  \dots & \dots & \dots              & \dots                 & \dots \\
MK421    &  0.20 &   2.7 &   3.0 &   2.8 & \dots & \dots & \dots \\
1156+295 &  2.34 &  10.7 &  28.3 &  47.5 & $  0.79 \pm  0.08 $ & $  0.105 \pm  0.019 $ & $3.5 \pm 1.8$ \\
1219+285 & 0.80  &   3.3 &  \dots & \dots & \dots              & \dots                 &  \dots \\
1222+216 &  0.98 &   1.8 &   5.6 &   5.4 & $  0.98 \pm  0.22 $ & $  0.222 \pm  0.009 $ & \dots & $\surd$ \\
3C273B   & 22.45 &  76.2 & 236.0 & 162.1 & $ 16.03 \pm  1.61 $ & $  0.174 \pm  0.016 $ & $-0.1 \pm 1.2$ \\
3C279    & 22.98 &  71.6 & 152.4 & 145.3 & $ 28.09 \pm  2.81 $ & $  0.342 \pm  0.016 $ & $17.5 \pm 0.5$ \\
1406-076 &  1.30 &   3.5 &  13.2 &  16.6 & $  1.77 \pm  0.20 $ & $  0.291 \pm  0.022 $ & $53.4 \pm 40.1$ & $\surd$ \\
1502+106 &  0.77 &   1.6 &   2.1 &   8.5 & $  0.77 \pm  0.08 $ & $  0.217 \pm  0.011 $ & \dots \\
1510-089 &  0.95 &   3.9 &  11.2 &  17.8 & $  1.17 \pm  0.13 $ & $  0.218 \pm  0.027 $ & $0.8 \pm 8.1$ \\
1606+106 &  1.07 &   3.4 &   5.0 &   6.5 & $  0.48 \pm  0.06 $ & $  0.199 \pm  0.016 $ & $177.3 \pm 56.7$ & $\surd$ \\
1633+38  &  2.07 &   2.8 &   7.0 &  34.1 & $  2.07 \pm  0.34 $ & $  0.113 \pm  0.011 $ &  $-21.2 \pm 7.4$ \\
1739+522 &  0.93 &   2.2 &   1.7 &   3.8 & $  0.93 \pm  0.27 $ & $  0.240 \pm  0.008 $ & \dots & $\surd$ \\
BLLAC   &  4.45 &   5.6 &  11.4 &  74.9 & $  0.72 \pm  0.09 $ & $  0.000 \pm  0.039 $ & $0.3 \pm 5.8$ \\
2209+236 &  1.02 &   2.1 &   2.1 &  10.5 & $  1.02 \pm  0.10 $ & $  0.193 \pm  0.007 $ & \dots & $\surd$ \\
CTA102   &  4.84 &  13.1 &  29.5 &  78.1 & $  1.77 \pm  0.18 $ & $  0.002 \pm  0.036 $ & $-2.3 \pm 3.3$ \\
3C454.3  &  5.62 &   5.7 &  26.4 &  56.1 & $  1.05 \pm  0.12 $ & $  0.001 \pm  0.033 $ & $22.4 \pm 5.3$ \\
\hline
\multicolumn{9}{c}{non-EGRET sources} \\
\hline
1413+135 &  2.16 &   7.9 &  22.9 &  32.0 & $  1.29 \pm  0.13 $ & $  0.076 \pm  0.031 $ & $-1.0 \pm 3.5$ & $\surd$ \\
1655+077 &  1.53 &   1.9 &   7.1 &  12.1 & $  1.53 \pm  0.10 $ & $  0.244 \pm  0.010 $ & \dots & $\surd$ \\
DA406    &  2.22 &   4.4 &   9.3 &  16.4 & $  0.45 \pm  0.05 $ & $  0.010 \pm  0.069 $ & $-5.1 \pm 6.3$ \\
NRAO512  &  0.79 &   1.8 &   2.0 &   5.6 & $  0.79 \pm  0.05 $ & $  0.204 \pm  0.008 $ & \dots & $\surd$ \\
\enddata
\end{deluxetable}

\begin{references}

\reference{hartm99} Hartman, R.C. \etal, 1999, \apjs, 123, 79

\reference{kelle98} Kellermann, K.I., Vermeulen, R.C., Zensus, J.A. \& Cohen, M.H., 1998, \aj, 115, 1295

\reference{lonsd98} Lonsdale, C., Doeleman, S. \& Phillips, R.B., 1998, \aj,
116, 8

\reference{matto97} Mattox, J.R., \etal, 1997, \apj, 481, 95

\reference{moell96} Moellenbrock, G.A., et al., 1996, \aj, 111, 2174

\reference{roger94a} Rogers, A.E.E., 1994a, {\it Workshop on the Coordination of
Millimeter VLBI}, Haystack Observatory

\end{references}
\end{document}